\documentclass[3p,times,procedia]{elsarticle}
\usepackage{nupha_ecrc}

\volume{00}

\firstpage{1}

\journalname{Nuclear Physics A}

\runauth{}

\jid{nupha}

\jnltitlelogo{Nuclear Physics A}

\usepackage{amssymb}

\usepackage[figuresright]{rotating}

 \newcommand{\kst}{$\mathrm{K}^{*0}$}
 \newcommand{\pt}{$p_\mathrm{T}$}
\newcommand{\mpt}{$\langle p_{\mathrm{T}} \rangle$}
\newcommand{\snn}{$\sqrt{s_{\mathrm{NN}}}$}

\begin{document}
\begin{frontmatter}



\dochead{XXVIIth International Conference on Ultrarelativistic Nucleus-Nucleus Collisions\\ (Quark Matter 2018)}

\title{Multiplicity dependence of strangeness and hadronic resonance production in pp and p-Pb collisions with ALICE at the LHC}
\author{Ajay Kumar Dash (For the ALICE collaboration)}
\address{School of Physical Sciences, National Institute of Science Education and Research, HBNI, Jatni-752050, India}
\begin{abstract}
One of the key results of the LHC Run 1 was the observation of an enhanced production of strange particles in high 
multiplicity pp and p--Pb collisions at \snn~= 7 and 5.02 TeV, respectively. The strangeness enhancement is investigated 
by measuring the evolution with multiplicity of single-strange and multi-strange baryon production relative to non-strange 
particles. A smooth increase of strange particle yields relative to the non-strange ones with event multiplicity has been 
observed in such systems. We report the latest results on multiplicity dependence of  strange and multi-strange hadron 
production in pp collisions at $\sqrt{s} = $ 13 TeV with ALICE. We also presented recent measurements of mesonic and 
baryonic resonances in small collision systems like pp and p--Pb at \snn~= 13 and 8.16 TeV, respectively. The system size 
dependent studies in pp and p-Pb collisions have been used to investigate how the hadronic scattering processes affect 
measured resonance yields and to better understand the interplay between canonical suppression and strangeness 
enhancement. The measurement of the $\phi(1020)$ meson as a function of multiplicity provides crucial constraints in 
this context. 
\end{abstract}
\end{frontmatter}
\section{Introduction}
\label{Intro}
The enhancement of strangeness production in high-energy nucleus-nucleus collisions (A--A) relative to the production 
in proton-proton (pp) collisions has historically been proposed as one of the signatures of Quark-Gluon Plasma (QGP) 
formation \cite{qgp}. Experimentally, this was first observed in Pb--Pb collisions at SPS \cite{spsqgp} and subsequently 
at RHIC \cite{rhicqgp} and LHC \cite{lhcqgp}. This  enhancement in A--A collisions has also been explained as a 
suppression of strangeness production in reference samples like pp and p--Pb due to the lack of phase space for strange 
quark creation in small systems \cite{cansup}. Recently, the ALICE Collaboration has observed that the yields of strange 
hadrons are enhanced relative to non-strange hadrons in high multiplicity pp collisions at $\sqrt{s} = $ 7 TeV and the 
strangeness enhancement in high multiplicity pp collisions reaches the values observed in p-Pb and peripheral Pb–Pb 
collisions \cite{alicenature}. The strength of this enhancement increases with increasing strange quark content of the hadron. 
It was observed that the enhancement of strangeness production in small systems is due to the strangeness content rather 
than mass or a meson-baryon effect. As the net strangeness of $\phi$-meson is zero, it is interesting to study how the 
$\phi$-meson behaves in the strangeness enhancement picture. It has been observed that by comparing similar event 
multiplicity and by changing the colliding system (pp, p--Pb and Pb--Pb) the relative particles abundances are not modified. 
By comparing the results from the pp collisions at $\sqrt{s} =$ 13 TeV to those at lower energies, the center-of-mass energy 
dependence of hadrochemistry can be isolated. It is also interesting to measure the short-lived resonance particles like 
\kst~($\tau\sim4.2$ fm/$c$) and $\phi$ ($\tau\sim46.3$ fm/$c$). These hadrons may decay during the hadronic phase 
(phase between chemical and kinetic freeze-out). As a result, the decay daughters may re-scatter, leading to a reduction 
in the measurable resonance yields. They may also be regenerated due to pseudo-elastic scattering of hadrons through 
a resonance state, which enhances their production \cite{regen}. The centrality or multiplicity dependent suppression of 
the \kst/K ratio has been observed in p--Pb and Pb--Pb collisions \cite{kstpPb, kstPbPb}. So, it is most important to see 
if such an effect can be observed in high-multiplicity pp collisions, which might be an indication for a hadronic phase with 
non-zero lifetime. In this work, we present the multiplicity dependence of resonance, strange and multi-strange hadron 
production in pp and p--Pb collisions measured with the ALICE detector at \snn~= 13 and 8.16 TeV, respectively.
\section{Analysis}
\label{rec}
A detailed description of the ALICE apparatus can be found in ref. \cite{aliceDet}. The main detectors which are relevant to 
this analysis are the Time Projection Chamber (TPC), the Time-of-Flight detector (TOF), the Inner tracking system (ITS) 
(covering the mid-rapidity window of $|\eta| < 0.9$) , and the V0 (V0A covering $2.8 < \eta < 5.1$ and V0C covering 
$-3.7 < \eta < -1.7$) detector. The multiplicity classes are defined based on percentiles of the distribution of the summed 
amplitudes measured in the V0 detectors (V0A+V0C) \cite{vocen}. The measurements of resonance, strange and multi-strange 
hadron production are performed at mid-rapidity ($|y| < 0.5$ in pp collisions and $0 < y_{cm} < 0.5$ in p-Pb collisions) as a 
function of the charged particle density, which is also measured at mid-rapidity for each multiplicity class. The aforementioned 
measurements are performed via the invariant mass analysis based on the following decay channels (branching ratios): 
\kst ($\bar{\mathrm{K^{*0}}}$)$\rightarrow$K$^{+}\pi^{-}$(K$^{-}\pi^{+}$) (66.6 \%), $\phi\rightarrow$K$^{+}$K$^{-}$ (48.2 \%),
K$_{S}^{0} \rightarrow$ $\pi^{+}\pi^{-}$ (69.2\%), $\Lambda \rightarrow$ p$\pi^{-}$ (63.9\%), $\Xi \rightarrow \Lambda\pi^{-}$ 
(99.9\%) and $\Omega \rightarrow \Lambda$K$^{-}$(67.8\%). In the invariant-mass method one needs to estimate the 
combinatorial background, which is evaluated by using an event-mixing technique for resonances. For strange and multi-strange 
hadrons, a set of topological cuts is applied to eliminate the candidates which do not fit the expected decay topology. After the 
background subtraction the raw yields are extracted from the signal distribution to be then corrected for detector acceptance, 
tracking efficiency and branching ratio. The first two corrections were determined by means of  Monte Carlo simulations of the 
ALICE detector response.
\section{Results and discussion}
\label{res}
\begin{figure}[h]
 \begin{minipage}[c]{0.65\textwidth}
\vspace{-0.4 cm}
\includegraphics[scale=0.47]{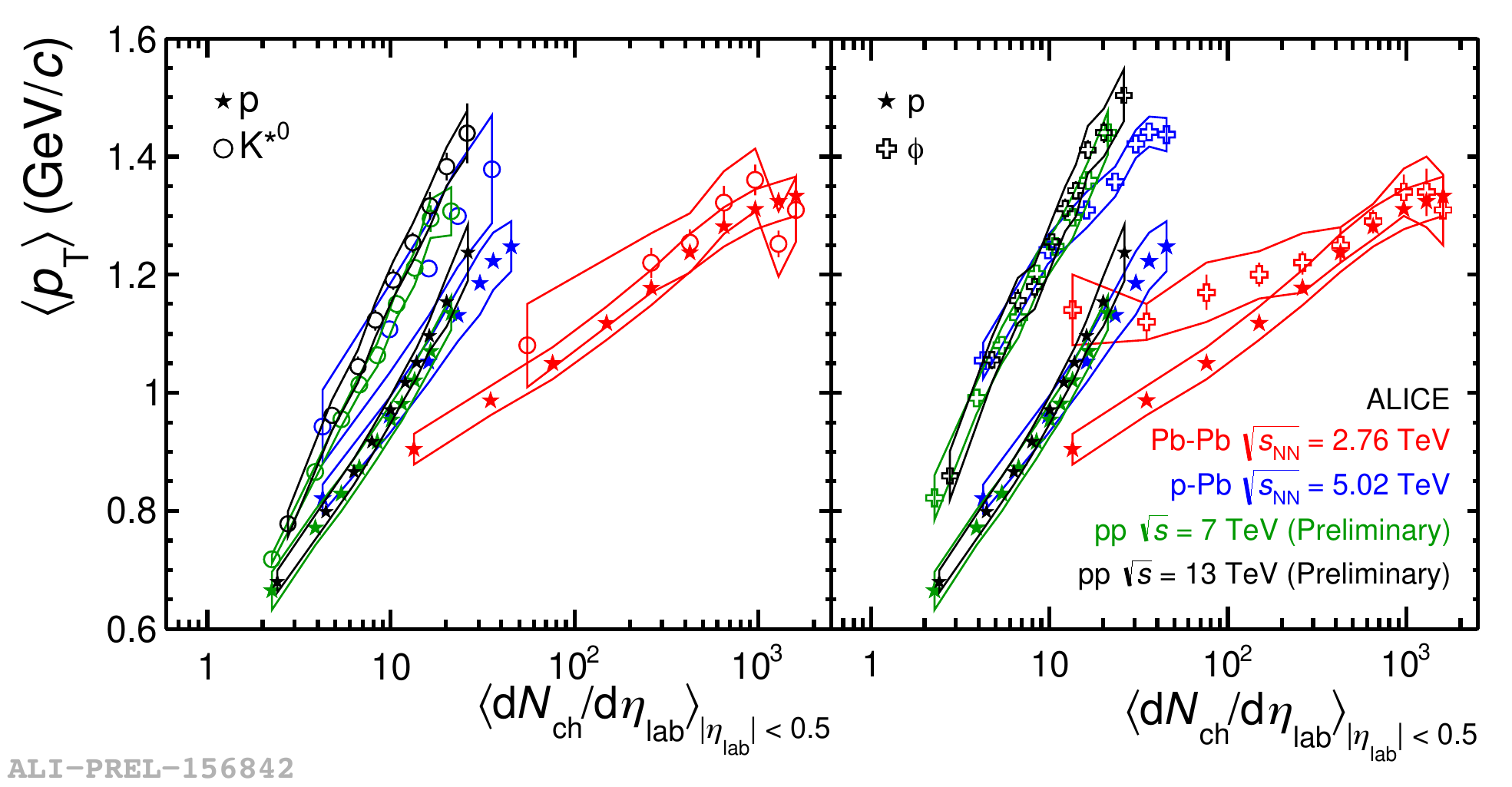}
\vspace{-0.4cm}
  \end{minipage}\hfill
  \begin{minipage}[c]{0.35\textwidth}
    \caption{(Color online) Mean transverse momentum (\mpt)~ of \kst, $\phi$ and p in pp (at $\sqrt{s}$ = 7 and 13 TeV), p--Pb 
    (at \snn~= 5.02 TeV) and Pb-Pb (at \snn~= 2.76 TeV) collisions as functions of multiplicity. The bars represent the statistical 
    error and the lines represents the systematic error.}
     \label{meanpt}
  \end{minipage}
\end{figure}
\begin{figure}[h]
  \begin{minipage}[c]{0.65\textwidth}
\includegraphics[scale=0.24]{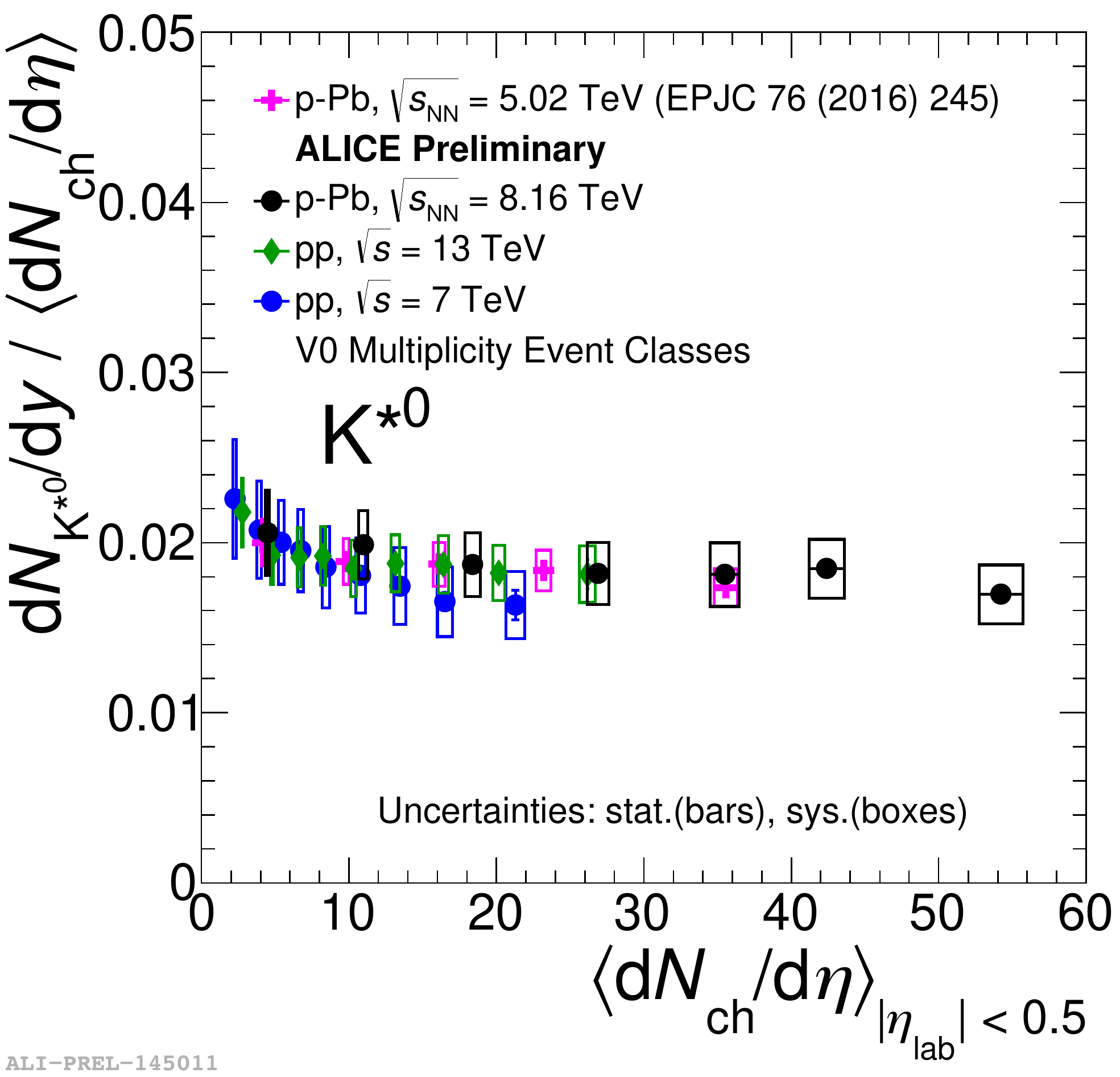}
\includegraphics[scale=0.24]{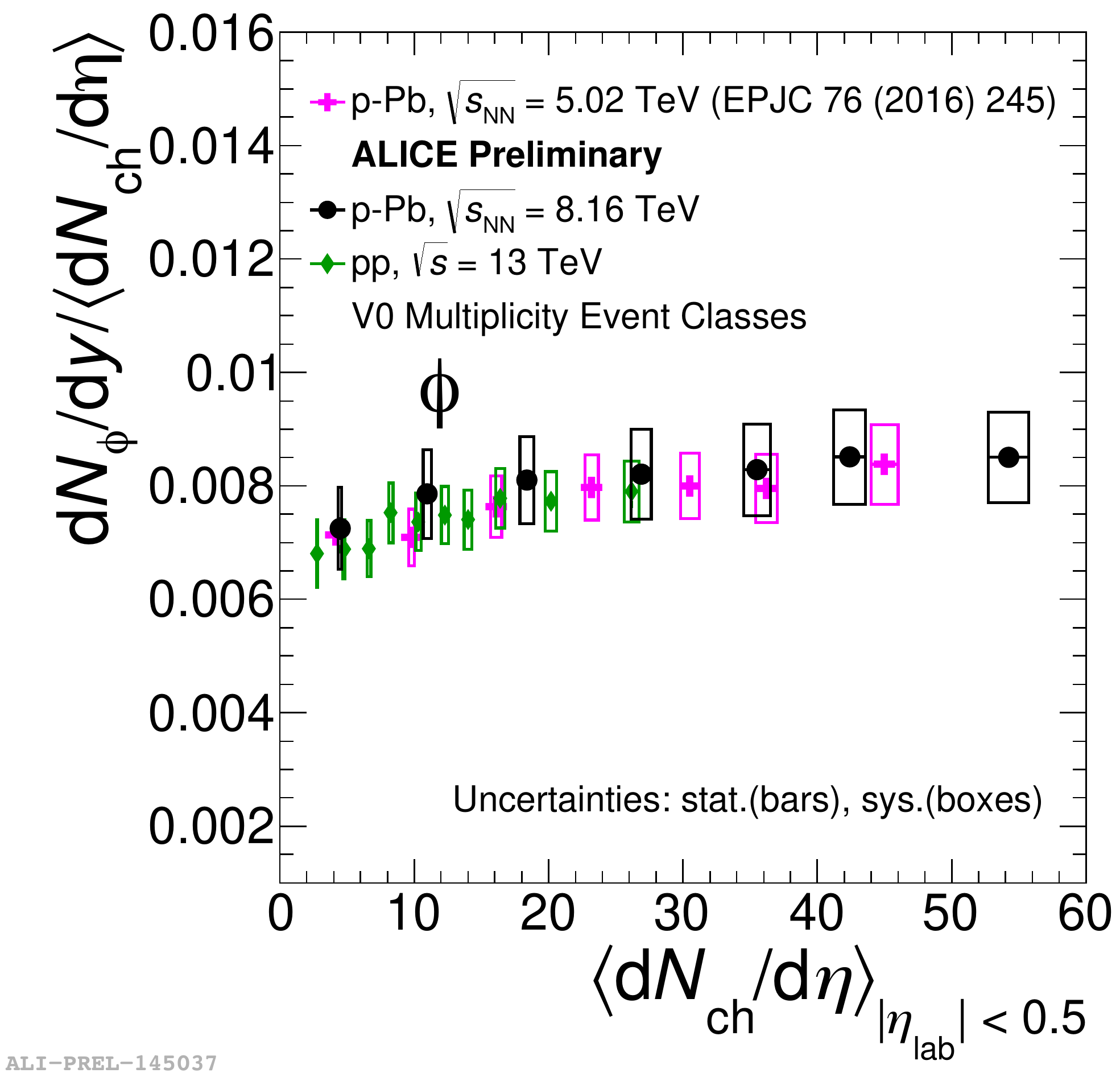}\\    
\vspace{-0.4cm}
  \end{minipage}\hfill
  \begin{minipage}[c]{0.32\textwidth}
    \caption{(Color online) Integrated yields of K$^{*0}$ (left panel)  and $\phi$ (right panel) normalized to 
    $\langle\mathrm{d}\it N_{\mathrm{ch}}/\mathrm{d}\eta\rangle$ in pp collisions (at $\sqrt{s}$ = 7 and 13 TeV) and  p--Pb 
    collisions (at \snn~= 5.02 and 8.16 TeV) for different multiplicity classes. The bars and the boxes represent the statistical 
    and systematic error, respectively.} \label{dndy}
  \end{minipage}
\end{figure}
\begin{figure}[h]
  \begin{minipage}[c]{0.65\textwidth}
  \vspace{-0.3cm}
\includegraphics[scale=0.25]{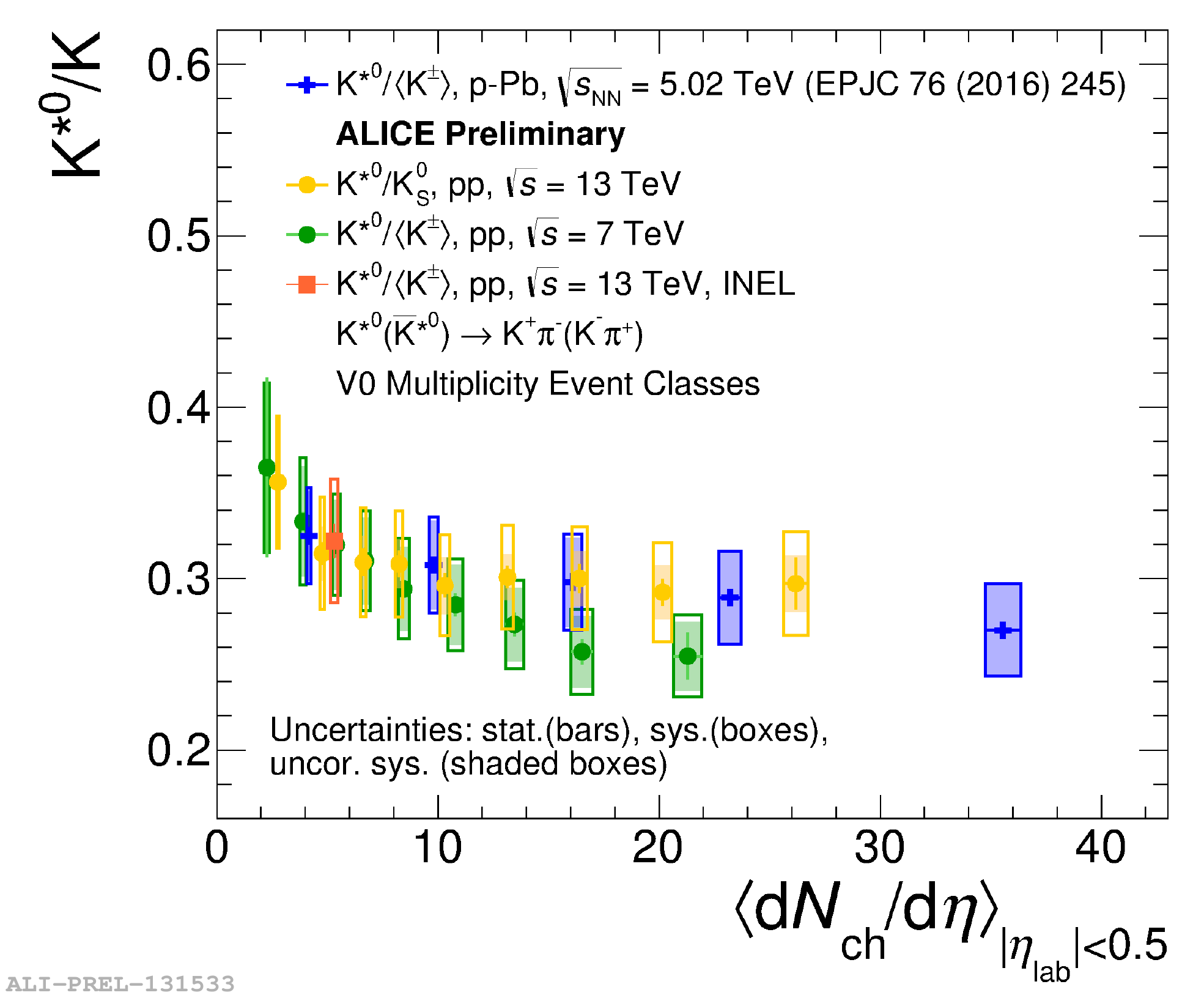}
  \includegraphics[scale=0.222]{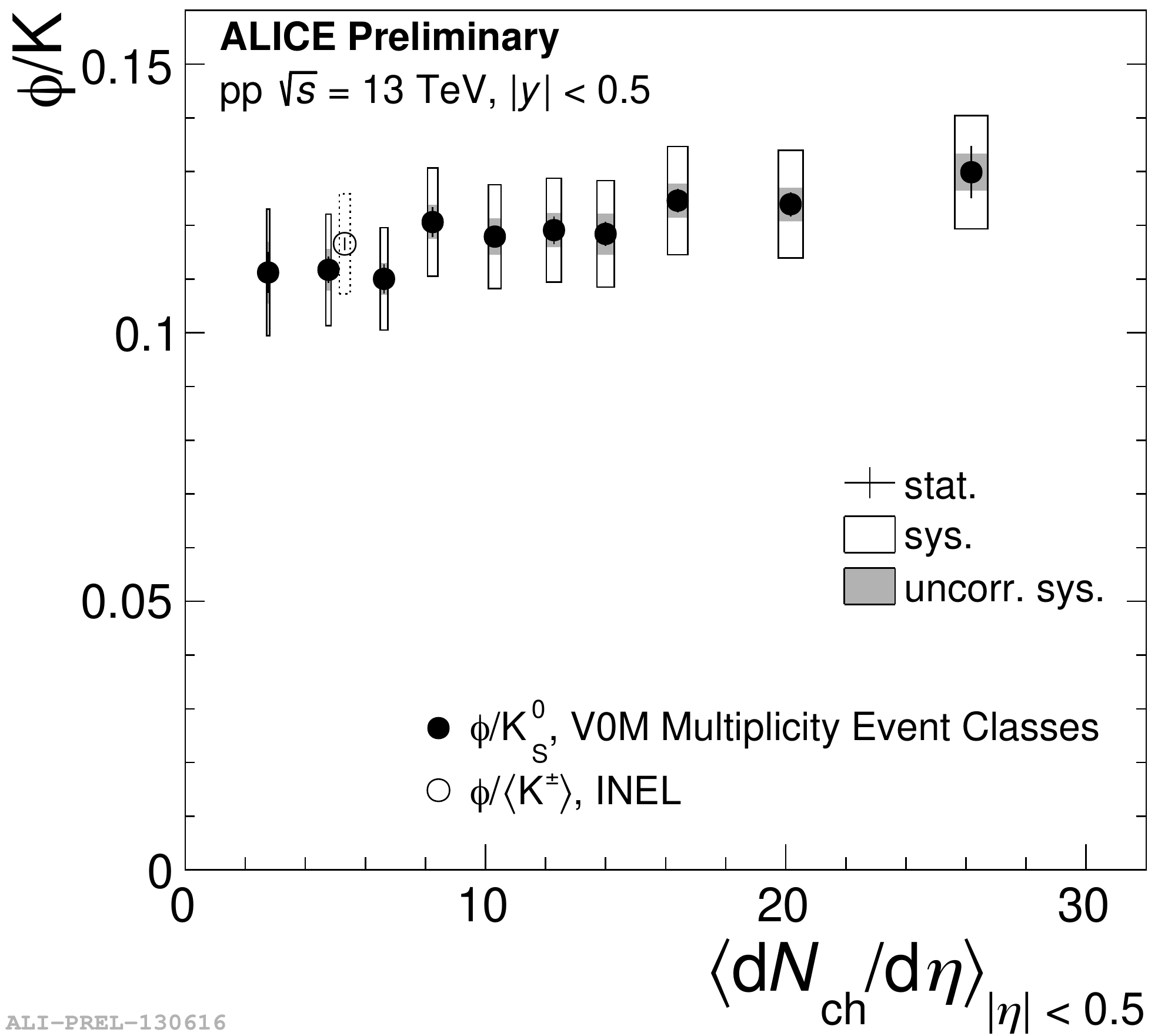}\\
  \vspace{-1.0cm}
  \end{minipage}\hfill
  \begin{minipage}[c]{0.32\textwidth}
    \caption{(Color online) Ratios of the integrated yields, K$^{*0}$/K (left panel) measured in pp collisions at $\sqrt{s}$ = 7 
    and 13 TeV and p--Pb collisions at \snn~= 5.02 and 8.16 TeV and $\phi$/K (right panel) in pp collisions at $\sqrt{s}$ = 13 
    TeV for different multiplicity classes. The bars and the boxes represent the statistical and systematic error, respectively.} 
    \label{res:ratio}
  \end{minipage}
\end{figure}
\begin{figure}
  \begin{minipage}[c]{0.6\textwidth}
  \vspace{-0.4cm}
    \includegraphics[scale=0.47]{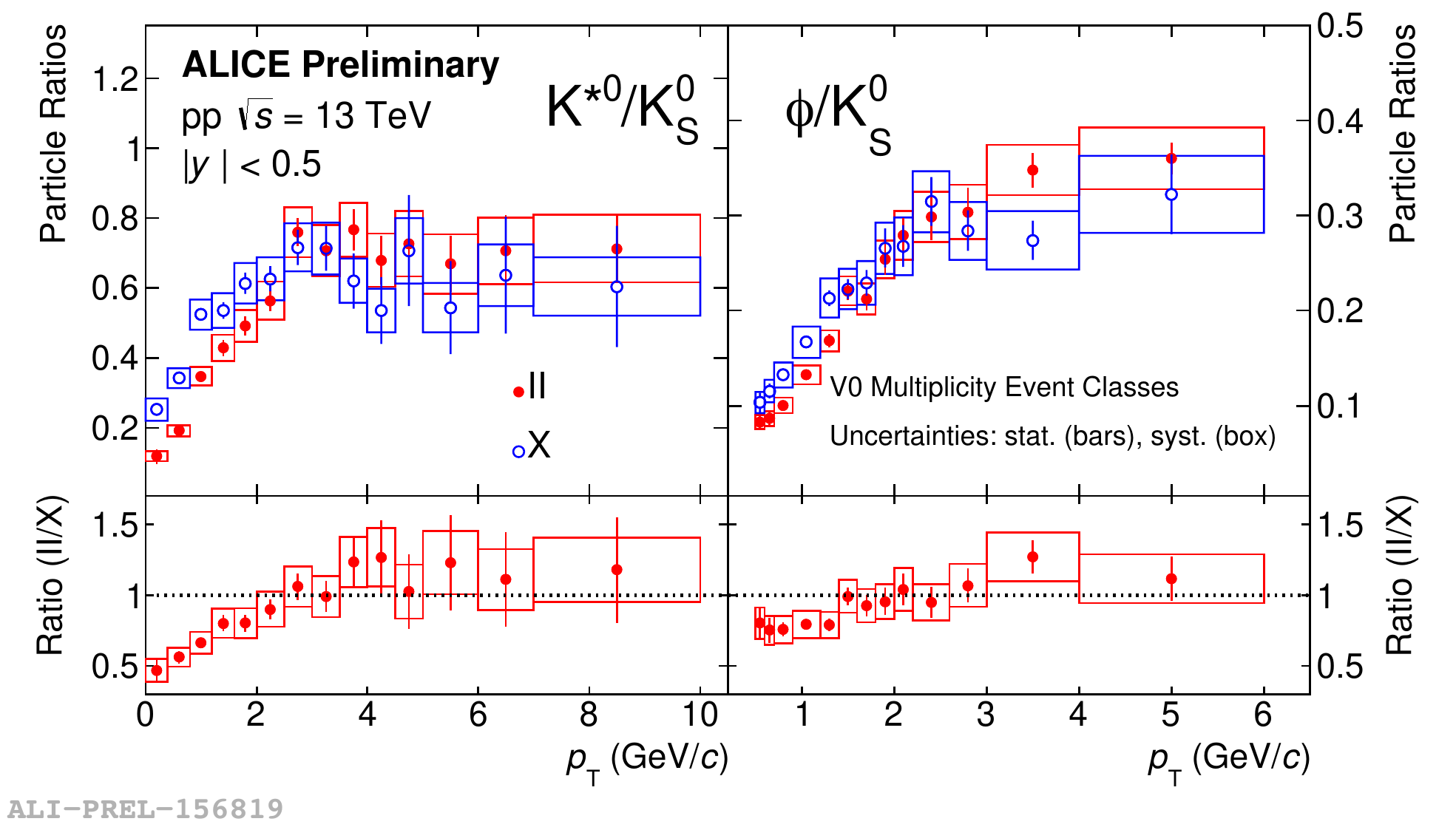}
    \vspace{-0.2cm}
  \end{minipage}\hfill
  \begin{minipage}[c]{0.35\textwidth}
    \caption{(Color online) \pt~differential ratio of  K$^{*0}$/K${_\mathrm{S}^0}$ (left panel) and $\phi$/${_\mathrm{S}^0}$ (right panel) in pp collisions at 
    $\sqrt{s}$ = 13 TeV for two different extreme multiplicity classes, where the $\langle\mathrm{d}\it N_{\mathrm{ch}}/\mathrm{d}\eta\rangle$ 
    in high (II) and low (X) multiplicity classes are $\sim$ 20 and 2.4, respectively. The bottom panel shows the ratio between the yield ratio 
    in high to low multiplicity class. The bars and the boxes represent the statistical and systematic error, respectively.} 
    \label{diff:resRatio}
  \end{minipage}
\end{figure}
The production of \kst, $\phi$, K$_{S}^{0}$, $\Lambda$, $\Xi$ and $\Omega$ is measured in different multiplicity classes in pp 
collisions at $\sqrt{s}$ = 13 TeV in a wide \pt~range. In addition, the resonances are also measured in several multiplicity classes 
in p--Pb collisions at \snn~= 8.16 TeV. The  $p_\mathrm{T}$ integrated  hadron yields ($\mathrm{d}\it{N}/\mathrm{d}\it{y}$) and 
mean $p_\mathrm{T}$ ($\langle p_{\mathrm{T}} \rangle$) for each multiplicity event class are determined by integrating the 
$p_\mathrm{T}$ spectra in the measured range and by using a  L\'{e}vy-Tsallis fit function to extrapolate the yields in the 
unmeasured \pt~region. Figure ~\ref{meanpt} shows the $\langle p_{\mathrm{T}} \rangle$ of $\mathrm{K}^{*0}$, $\phi$ 
and proton as a function of the average charged particle multiplicity density ($\langle\mathrm{d}\it N_{\mathrm{ch}}/\mathrm{d}\eta\rangle$) 
measured at mid-rapidity ($|y| < 0.5$) in pp collisions at $\sqrt{s}$ = 13 TeV and compared with the results obtained in pp, p--Pb and Pb--Pb 
collisions at \snn~= 7, 5.02  and 2.76 TeV, respectively. For all the particles studied an increase in $\langle p_{\mathrm{T}} \rangle$ 
from low to high multiplicity classes is observed. The same increasing  trend of the \mpt~as a function of the multiplicity is observed 
in pp collisions at $\sqrt{s}$ = 7 TeV and 13 TeV and a mass ordering of \mpt~is found to be followed in central and semi-central 
Pb--Pb collisions, i.e., particles with similar masses have similar \mpt~values, as expected from the hydrodynamic expansion of the 
system \cite{hydro}. However, this breaks down for smaller systems. The increase in \mpt~is steeper for smaller systems. The yields 
normalized to the $\langle\mathrm{d}\it N_{\mathrm{ch}}/\mathrm{d}\eta\rangle$ of \kst and $\phi$ in pp collisions at $\sqrt{s}$ = 7 
and 13 TeV, p-Pb collisions at \snn~= 5.02 and 8.16 TeV are shown as a function of multiplicity in Fig. \ref{dndy}. It is observed that 
at similar multiplicity, particle production is independent of the system size and collision energy. The yield ratios of resonances to long 
lived hadrons as a function of $\langle\mathrm{d}\it N_{\mathrm{ch}}/\mathrm{d}\eta\rangle$ for pp collisions at $\sqrt{s}$ = 7 and 13 TeV 
and p--Pb collisions at \snn~= 5.02 TeV are shown in Fig. \ref{res:ratio}. The ratio $\phi$/K , $\Sigma^{*\pm}/\Lambda$ (not shown here) 
and $\Lambda^{*}/\Lambda$ (not shown here) are independent of the event multiplicity in pp and p-Pb, consistent with 
re-scattering/re-generation effects. A hint of a decrease in the \kst/K${_\mathrm{S}^0}$ ratio is seen from low to high multiplicity pp and 
p--Pb collisons, which might be due to re-scattering. As re-scattering is important at low \pt, we looked into the \pt~differential 
ratios \kst/K${_\mathrm{S}^0}$ and $\phi$/K${_\mathrm{S}^0}$, which are shown in Fig.~\ref{diff:resRatio} for two extreme multiplicity classes, 
\kst/K${_\mathrm{S}^0}$ in the left panel and $\phi$/K${_\mathrm{S}^0}$ in the right panel. In the bottom panels the ratios of high to low multiplicity 
classes are shown. Both ratios increase at low \pt~and saturate for \pt ~$>$ 3 GeV/$\it{c}$. Depletion in the ratios for high multiplicity 
at low \pt~is observed. From the bottom panels it is clear that the \kst/K${_\mathrm{S}^0}$ ratio is more suppressed compared to $\phi$/K${_\mathrm{S}^0}$ 
for \pt $~<$ 2 GeV/$\it{c}$. This observation is consistent with a re-scattering effect in high multiplicity pp collisions.
\begin{figure}
  \begin{minipage}[c]{0.65\textwidth}
  \includegraphics[scale=0.22]{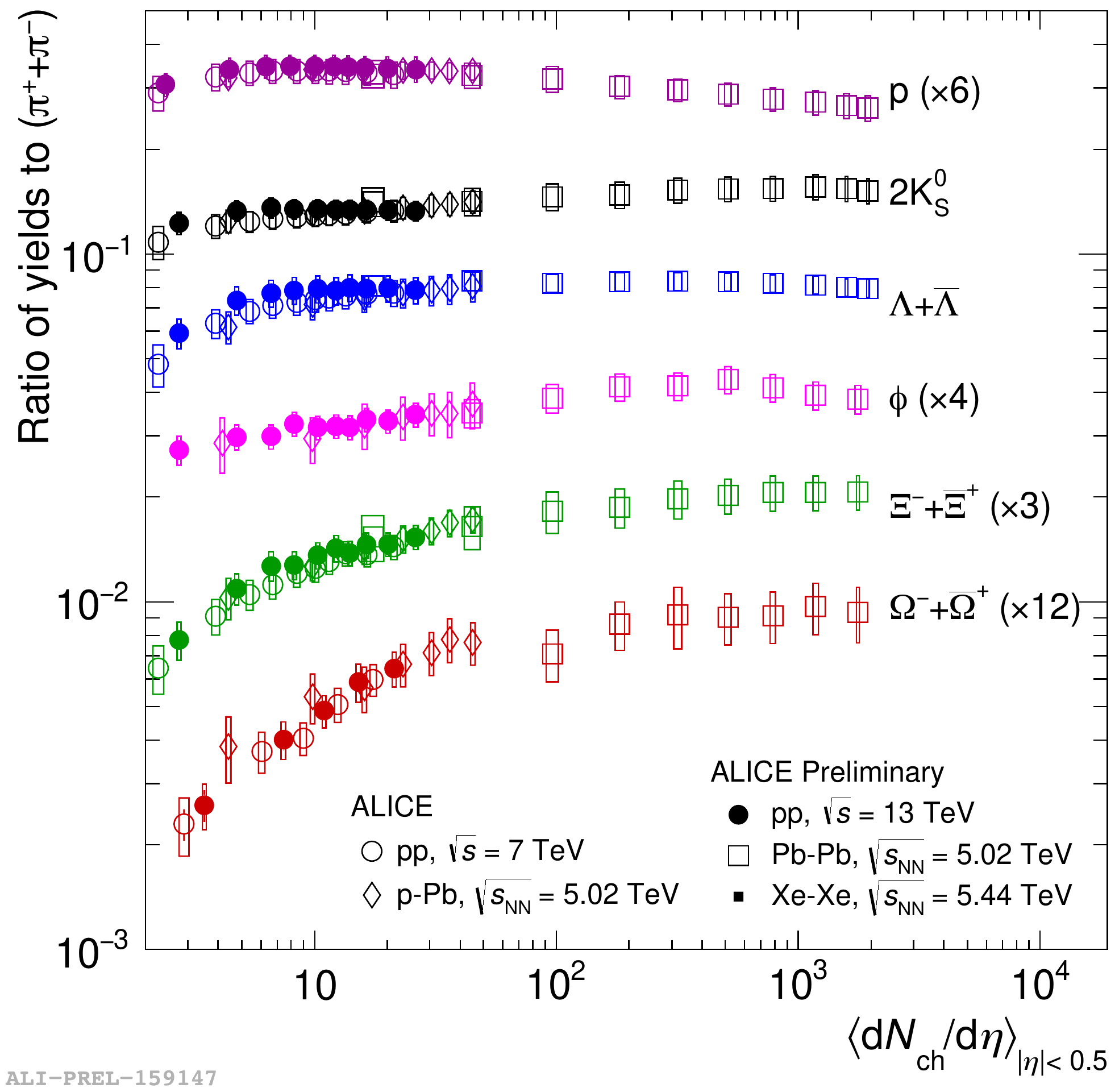}
\includegraphics[scale=0.24]{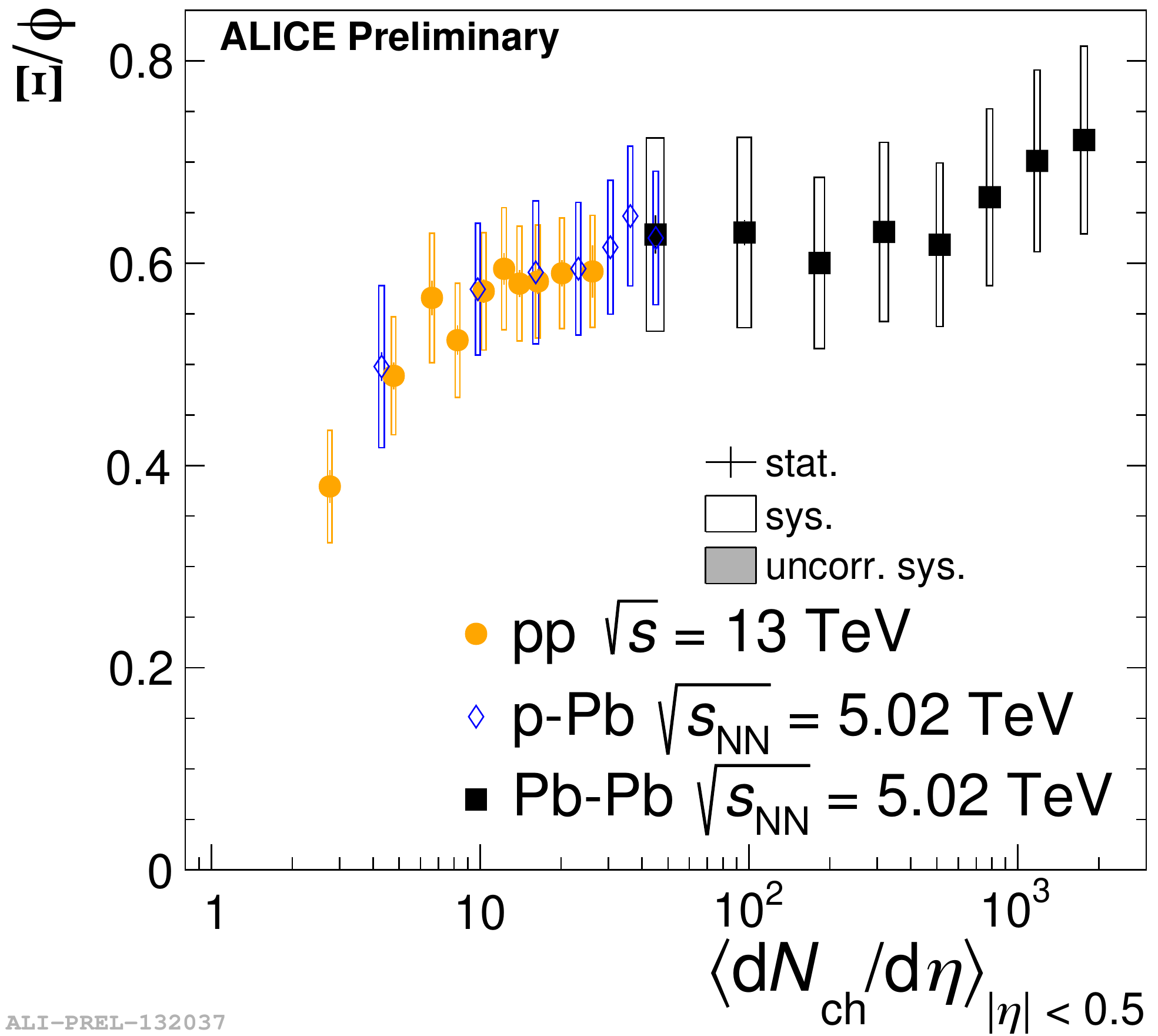}
\vspace{-1.0cm}
  \end{minipage}\hfill
  \begin{minipage}[c]{0.35\textwidth}
    \caption{(Color online) Left panel: Ratios of the integrated yields, (p+$\bar{\mathrm{p}}$)/($\pi^{+}+\pi^{-}$), 2K$_\mathrm{S}^{0}$/($\pi^{+}+\pi^{-}$), 
    ($\Lambda+\bar{\Lambda}$)/($\pi^{+}+\pi^{-}$), $2\phi$/($\pi^{+}+\pi^{-}$),  ($\Xi+\bar{\Xi}$)/($\pi^{+}+\pi^{-}$), 
    ($\Omega+\bar{\Omega}$)/($\pi^{+}+\pi^{-}$), measured as a function of the charged particle density in pp 
    (at $\sqrt{s}$ = 7 and 13 TeV), p--Pb (at \snn~= 5.02 TeV) and Pb-Pb (at \snn~= 5.02 TeV) collisions. Right panel: 
    Ratios of the integrated yields, ($\Xi+\bar{\Xi}$)/$\phi$ in pp collisions at $\sqrt{s}$ = 13 TeV, p--Pb and Pb-Pb 
    collisions at \snn~= 5.02 TeV. The bars and the boxes represent the statistical and systematic error, respectively.} 
    \label{stratio}
  \end{minipage}
\end{figure}
The ratio of particle yields to pions as a function of $\langle\mathrm{d}\it N_{\mathrm{ch}}/\mathrm{d}\eta\rangle$ in pp at 
$\sqrt{s}$ = 7 and 13 TeV, p--Pb and Pb-Pb at \snn~= 5.02 TeV is shown in the left panel of Fig.~\ref{stratio}. The ratios 
from pp collisions at $\sqrt{s}$ = 13 TeV are in agreement with the measurement at lower energy. A smooth transition 
is observed between pp, p-Pb and Pb-Pb: at similar multiplicity strangeness production is independent of collision system 
and energy. The $\phi$ also shows strangeness enhancement from low to high multiplicity. The ratios $\phi$/K${_\mathrm{S}^0}$ and 
$\Xi/\phi$ (shown in the right panel of Fig.~\ref{stratio}  for pp collisions at $\sqrt{s}$ = 13 TeV and p--Pb and Pb--Pb collisions 
at \snn~= 5.02 TeV) show a flat or slowly increasing trend over a wide range of multiplicity. This suggests that the $\phi$ 
behaves like a particle with effective strangeness between 1 and 2, similar to K and $\Xi$. This is important, since $\phi$ 
is not subject to canonical suppression, whereas K and $\Xi$ should be canonically suppressed.
\section{Summary}
The ALICE collaboration reported results on resonance, strange and multi-strange particle production as a function of 
multiplicity in pp and p--Pb collisions at \snn~= 13 and 8.16 TeV, respectively. Measurements of particle spectra at 
different energies as a function of multiplicity indicate that the hadrochemistry is rather driven by event multiplicity 
than by system size to collision energy. A hint of a decrease is observed for the \pt-integrated  \kst/K${_\mathrm{S}^0}$ ratio 
from low to high multiplicity pp and pPb collisions, while $\phi$/K${_\mathrm{S}^0}$ ratio remains constant. Those observations 
suggest the presence of re-scattering effects in high multiplicity pp collisions. Like other strange hadrons, the $\phi$ 
shows strangeness enhancement from low to high multiplicity class in pp collisions and behaves like a particles 
with net strangeness between 1 and 2. The strangeness enhancement depends on the strangeness content of 
the particle.
\label{sum}


 
\end{document}